# Astrophysical chaotic gun effect

Gheorghe Dumitrescu

*High School Toma N.Socolescu, street Gh.Gr.Cantacuzino No. 328*
*Ploiesti,Romania*
*meditatie@yahoo.com*

**Abstract.** We propose a kinetic equation for a special kind of acceleration: chaotic gun effect. Then we infer a distribution function which can depict the instability condition. With this distribution function we derive the power spectrum of the synchrotron emission and we prove the power law form of the power spectrum. We show that the spectral index of the emission spectrum is related to the spectral index of the number of the charged particles in the beam. Our numeric simulations show that the spectrum has a break at a frequency threshold where the chaotic acceleration becomes efficient. Assuming this threshold to the set on of the efficient chaotic gun effect we estimate the magnetic strength .Our paper advocates an electromagnetic process able to accelerate charged particles to high energies starting from low energies. Assuming the high-energy particles spectra of Mkn 501 to be produced by the synchrotron emission during chaotic gun effect we estimate some parameters of the source.

## 1.INTRODUCTION

Acceleration of charged particles by electromagnetic fields leads to the extraction of energy from these fields. In a stochastic scheme the acceleration element may be represented by a resonant wave-particle interaction within the framework of a (plasma) wave turbulence. The Comptonization of ubiquitous cosmic (real) photons can be realized by a second order Fermi (stochastic) acceleration mechanism that extracts energy from electromagnetic fields. The nonlinear interaction between the relativistic charged cosmic particles and the composite field (cosmic photons and vehicle static magnetic field) represents the starting point for the present paper. Medvedev 2000[1] assumed that the electron acceleration takes place in the shock front of plasma in his model concerning small-angle "jitter" radiation . The earlier papers concerning the acceleration on small-scale random magnetic fields [2],[3]and [4] emphasized isotropic random magnetic field which leads to the so called diffusive shock acceleration mechanism. The fundamental assumption of the theory of diffusive shock acceleration is that accelerated particles diffuse in space, i.e. that the particle flux is proportional to the gradient of the particle density (Fick's law). Charged particles deflected by fluctuations in the electromagnetic fields obey this relation only if their velocities are distributed almost isotropically. At a shock front, the downstream plasma speed is of the same order as the thermal speed of the ions in the plasma, so as result the theory of acceleration across the shock does not apply to particles whose energy is several times lower than the thermal energy. The question of how particles might be accelerated from the thermal pool up to an energy where they can be assumed to diffuse is referred to as the 'injection problem', and cannot be treated within the framework of the diffusive acceleration theory[5],[6]). L. O'C. Drury et al.2001[7] suggest that waves are excited by collective instability of ions reflected from a perpendicular shock, and that these waves damp on thermal electrons, thereby accelerating them. Such a process has been proposed by Galeev (1984)[8] as a possible acceleration mechanism for cosmic ray electrons. Instabilities driven by shock–reflected ions at SNR shocks have also been invoked by Papadopoulos (1988)[9] and Cargill and Papadopoulos (1988)[10] as mechanisms for electron heating, rather than electron acceleration.

Medvedev and Loeb(1999) suggest that the field might originate from a highly magnetized stellar remnant, such as a neutron star, with B~$10^{16}$ G and a turbulent magnetic dynamo could amplify a relatively weak seed magnetic field in the vicinity of the progenitor. This process requires the turbulence to be anisotropic [11].

The need for a practical solution of the acceleration problem in the non-linear regime has been recognized by Berezhko & Ellison (1999)[12] . Some promising analytical solutions of the problem of non-linear shock acceleration have appeared in literature[13],[14].

An earlier and promising numerical experiment assuming a charged particle moving in a stochastic electromagnetic field revealed an explosive behavior of the chaotic motion: the charged particle is suddenly expelled from the region where it was accelerated [15]. This is the chaotic gun effect with which we will deal in our work. Its analytical approach emphasizes a random electromagnetic field superimposed on a uniform field. Any choice is possible for the type of the random field in the model adopted by Argyris and Ciubotariu [15].

Previously Hall and Sturrock(1967[16]) used the test-particle formulation of the problem of the behavior of charged particle in turbulent electromagnetic fields to derive a diffusion equation involving only second-order correlation functions of the field.

But chaotic gun effect leads to strong anisotropization of the distribution of charged particles in the direction of the electromagnetic wave vector. When anisotropization is very large the diffusion approximation breaks down. That is why one must to use a kinetic equation that keeps the time collision integral in the right hand. We propose an analytic mechanism concerning both anisotropic electron pitch angle distribution and synchrotron self-absorption starting from the numerical experiment of Argyris and Ciubotariu 2000[15].

In the second chapter of our paper we describe our model based on the numerical experiment of Argyris and Ciubotariu. We complete their approach in order to study the collective behavior of the charged particles contained in the beam. In the third chapter we derive a kinetic equation of an ensemble of charged particles in a random electromagnetic field and thereafter, from this equation, we find out the distribution function $f$.

Using the distribution function we infer in the fourth chapter the intrinsic spectrum of the synchrotron emission for the spectral index of the number of the particles, 7. Those indices that are natural numbers favor the analytical integration of the power of the emission. We show that at high energies the spectrum is a power law.

In the fifth chapter we point out some astrophysical consequences of our approach.

## 2.The model

A relativistic beam of charged particles gyrates in a uniform magnetic. We choose a reference frame where the uniform magnetic field $B_0$ is along z-axes. The bulk of the charged particles initially has such energies as the gyrofrequency ratio $\beta = \dfrac{|\Omega_b|}{\Omega_B}$ of the dimensionless frequencies $\Omega_b = \dfrac{\omega_b}{kc}$ of the wave emitted by the particles of the beam, and $\Omega_B = \dfrac{\omega_B}{kc}$ of the gyration in the uniform magnetic field is smaller than 4. Here $\omega_b$ is the frequency of the wave emitted by charged particles of the beam and $\omega_B$ is the nonrelativistic gyrofrequency of a particle.

But a charged particle moving at an angle to an external magnetic field behaves as a nonlinear oscillator which generates an electromagnetic field. The beam posses a symmetry such that it generates a perturbation of an electromagnetic field with a harmonic space-dependence and, generally, non-harmonic time-dependence of the form

$$\mathbf{E_b}(x,t) = \{E_{bx}, E_{by}, E_{bz}\} = \{\mathbf{Re}[E_x(t)e^{ikx}], \mathbf{Re}[E_y(t)e^{ikx}], \quad (2.1)$$
$$\mathbf{B_b}(x,t) = \{B_{bx}, B_{by}, B_{bz}\} = \{0, 0, \mathbf{Re}[B_b(t)e^{ikx}]\} \quad (2.2)$$

where

$$\mathbf{k} = \{k_x, k_y, k_z\} = \{k, 0, 0\} \quad (2.3)$$

is the corresponding wave vector and **Re** applied to a quantity refers to the real part of that quantity and $\mathbf{E_b}$ and $\mathbf{B_b}$ are the electric and magnetic field intensities generated by the beam.

Then the beam penetrates in a domain of the magnetic field where the strength has such magnitude as the ratio $\beta = \dfrac{H}{\Omega_B} = \dfrac{|\Omega_b|}{\Omega_B}$ that is larger than 4, which is the critical value of the onset of a new kind of acceleration. This can occur when two populations of charged particles emit at characteristic frequencies that are completely different (e.g. radio and gamma). For those particles whose characteristic frequency is much larger than of the other ones, the field emitted by the latter is the uniform magnetic field.

In their analytical model Argyris and Ciubotariu 2000[15] suppressed one of the electric components of $\mathbf{E_b}$, namely $E_{bx}$ from the equation (2.1) and assumed that the plasma frequency vanishes. In our model we will keep both components of the electric field but we will allow them to differ by their correlation lengths and spectral indices of the power of the emission Assuming a likely couple of charged particle populations mentioned above we will adopt the same value of the dimensionless plasma frequency as Argyris and Ciubotariu[15] did, namely $\Omega_p = 0$.

Since the equations that describe the motion were nonlinear Argyris and Ciubotariu performed a numerical experiment using a fifth order Runge-Kutta-algorithm with an adaptive step size control. The nonlinear equations generate a flow in a 3-dimensional phase space $(P_x, P_y, X)$. Numerical solutions showed that a strong acceleration emerges when the frequency $H$ of the wave produced by the beam (and where every particle is immersed) are larger than $0.6kc$. Here $k$ is the magnitude of the wavelength of wave emitted by the beam and $c$ is the speed of the light. Also every particle that meets the requirement mentioned above is suddenly expulsed from the domain where it is accelerated [fig.2.1 and fig.2.2].

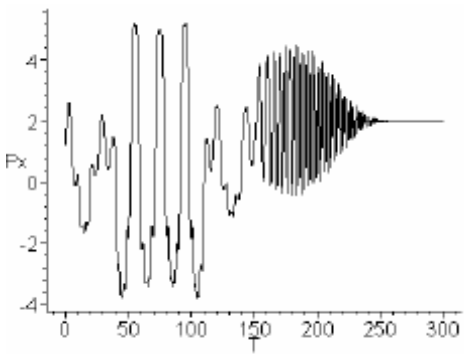
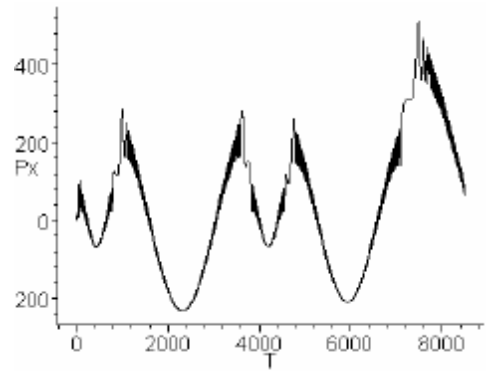

Fig.2.1 ($P_x$, $T$) Time series corresponding to the nonlinear equations of motion (a "flow" for $H=0.9$). A classical Fermi acceleration ("Chaotic or stochastic oscillations" along x- Axis) combined with the gun effect. Initial conditions: ($T, P_x, P_y, X$) = (0.0 0 1, 1. 5, 1.5, 0.001)

Fig.2 .2 ($P_x$, $T$) Time series for the case of a multi- gun cascade effects at some moments $T$ (e.g. the moment $T$; 900 ), the accelerated charged particle (large $P_x$) may be expulsed from combustion chamber in order to increase the speed of cosmic vehicle

Numerical simulations showed that near the resonances between cyclotron frequency harmonics and wave frequency

$$\omega = n\omega_c \quad \text{or} \quad n\frac{\Omega_B}{\gamma_n} = \frac{\omega}{kc} = 1 \quad (2.4)$$

a gun effect erupts, i.e. the charged particle is suddenly expulsed to a trajectory with larger radii. Here $n$ is the order of the resonance, $\omega_c$ is the relativistic gyrofrequency and $\gamma_n$ is the Lorentz factor corresponding to the n'th harmonic. Before the expulsion charged particle can extract energy from the magnetic field either as isolated resonance or overlapping resonances of magnetic field and gyration.

Before the expulsion the charged particle obeys a chaotic motion due to the high frequency oscillation with chaotic modulation of the amplitude. The localized chaotic regime is named "chaoson". On a specific Larmor spiral the particle goes around repeatedly, receiving a kick of energy each time it completes an orbit. We assume in our approach the synchrotron emission of beam to be the stochastic field responsible for chaotic acceleration.

While the x component of the momentum $p_x$ increases during multi-gun cascade, the y component $p_y$ is reduced to the same initial value after each stop of the multi-gun cascade.

If one increases the number of the iterations and the duration of the operation one observes a cascade of flow of "chaosons" with increasing energy.

The phase-portrait on the $(p_x, p_y)$ phase-space performed in the numerical experiment displayed a preferential direction of the acceleration, namely the electromagnetic wave vector direction.

In the following we will address our investigation concerning the collective acceleration only to the motion performed in a plane perpendicular to the uniform magnetic field, that is along x and y axis.

### 3. Kinetic equation for charged particles in stochastic and anysotropic electromagnetic fields

#### 3.1. The averaged kinetic equation

The problem of stochastic acceleration by a weak random electromagnetic field could be discussed by means of the Fokker-Planck equation. The coefficients in this equation can be expressed in terms of the turbulence spectrum. In this formulation of the problem, the interaction of particles by collisions and by collective interaction is neglected. The dissipation of the turbulence spectrum is also neglected.

Toptygin and Fleishman (1987[3]) inferred a kinetic equation for a relativistic particle in random electromagnetic field from the Boltzmann equation for the distribution function $f(\mathbf{r}, \mathbf{p}, t)$

$$\frac{\partial f}{\partial t} + \mathbf{v}\frac{\partial f}{\partial \mathbf{r}} + \mathbf{F}(\mathbf{r}, t)\frac{\partial f}{\partial \mathbf{p}} = 0 \quad (3.1.1)$$

where $\mathbf{F}(\mathbf{r},t) = q\,\mathbf{E}_b + \dfrac{q}{m_0 c\gamma}[\mathbf{p},(\mathbf{B}_0+\mathbf{B}_b)]$ is the total force acting on the charged particle. We will apply (3.1.1) in the reference frame of the motionless observer.

According to the chaotic gun effect model, the electric field has no regular component but the stochastic one, i.e. $\mathbf{E}_b$. One can also see that we identified $\mathbf{B}_b$ as the stochastic component of the magnetic field.

The Boltzmann equation (3.1.1) can be averaged by the use of the method of Vedenov et al. [4]. Averaging the equation (3.1.1) one finds the kinetic equation

$$\frac{\partial \tilde{f}}{\partial t} + \mathbf{v}\frac{\partial \tilde{f}}{\partial \mathbf{r}} - (\mathbf{\Omega}\partial)\tilde{f} = \int_0^\infty d\tau \left\{ \left(\frac{ec}{\varepsilon}\right)^2 \partial_\alpha T_{\alpha\beta}(\Delta\mathbf{r}(\tau),\tau)\partial_\beta + \right.$$

$$e^2 (\frac{\partial}{\partial p_\alpha}) K_{\alpha\beta}(\Delta\mathbf{r}(\tau),\tau)\frac{\partial}{\partial p_\beta} - \frac{e^2 c}{\varepsilon}\left(\partial_\beta S_{\alpha\beta}\frac{\partial}{\partial p_\alpha} + \frac{\partial}{\partial p_\alpha} S_{\alpha\beta}\partial_\beta\right) \Big\} \times$$

$$\tilde{f}(\mathbf{r}-\Delta\mathbf{r}(\tau),\mathbf{p}-\Delta\mathbf{p}(\tau),t-\tau) \qquad (3.1.2)$$

where

$$T_{\alpha\beta}(\mathbf{r},t) = \langle B_\alpha^{st}(\mathbf{r}_1,t_1) B_\beta^{st}(\mathbf{r}_2,t_2)\rangle, \quad \mathbf{r} = \mathbf{r}_1 - \mathbf{r}_2 \qquad (3.1.3)$$

$$K_{\alpha\beta}(\mathbf{r},t) = \langle E_\alpha^{st}(\mathbf{r}_1,t_1) E_\beta^{st}(\mathbf{r}_2,t_2)\rangle, \quad t = t_1 - t_2 \qquad (3.1.4)$$

$$S_{\alpha\beta}(\mathbf{r},t) = \langle E_\alpha^{st}(\mathbf{r}_1,t_1) B_\beta^{st}(\mathbf{r}_2,t_2)\rangle \qquad (3.1.5)$$

are the correlation tensors of the second order, and

$$\Omega = \frac{e\mathbf{B}_0 c}{\varepsilon} \qquad (3.1.6)$$

$$\frac{e}{c}[\mathbf{v},\frac{\partial}{\partial \mathbf{p}}] = \frac{ec}{\varepsilon}\partial \qquad (3.1.7)$$

For a stationary flow the correlation tensors used in the equation (3.1.2) depend on spatial coordinates and time as it follows

$$T_{\alpha\beta}(\mathbf{r}_1,t_1;\mathbf{r}_2,t_2) = T_{\alpha\beta}(\mathbf{r},t;\mathbf{x},\tau) \qquad (3.1.8)$$

where $\mathbf{r} = (\mathbf{r}_1 + \mathbf{r}_2)/2$; $\mathbf{x} = \mathbf{r}_1 - \mathbf{r}_2$; $t = (t_1 + t_2)/2$ and $\tau = t_1 - t_2$. The last argument $\tau$ allows for correlation weakening due to intrinsic motions of magnetic pulsations (e.g., waves with random phases). If $R_0 \gg L_0$, where $R_0$ is the radius of the particle motion in the magnetic field $B_0$ and $L_0$ is the correlation length of the small inhomogeneities then the momentum varies only slightly along the correlation length. Putting

$$\Delta\mathbf{p}(\tau) = 0, \Delta\mathbf{r}(\tau) = \mathbf{v}\tau \qquad (3.1.9)$$

in the equation (3.1.2) we come to the equation which describes the scattering of particles by small-scale inhomogeneities.

As the numerical simulations of the chaotic gun effect have displayed, at every complet gyration of the charged particle, the particle recieves a kick of energy when it reaches the wave vector direction. Hence the particle is deflected strictly along the x direction perpendicular to the uniform magnetic field $\mathbf{B}_0$. The acceleration takes place along the wave vector direction and there is no angular deflection. That is why we will omit in the equation (3.1.2) all the terms containing the $\grave{\mathbf{o}}$ operator. The $\grave{\mathbf{o}}$ operator is the angular variation of the velocity direction. The equation (3.1.2) becomes simpler

$$\partial \widetilde{f}/\partial t + \mathbf{v}(\partial \widetilde{f}/\partial \mathbf{r}) =$$

$$\int_0^\infty d\tau \left\{ e^2 (\frac{\partial}{\partial p_\alpha}) K_{\alpha\beta}(\Delta \mathbf{r}(\tau),\tau) \frac{\partial}{\partial p_\beta} \right\} \widetilde{f}(\mathbf{r}-\Delta \mathbf{r}(\tau),\mathbf{p},t-\tau) \qquad (3.1.10)$$

For correct description of the particle emission at $\omega >> \omega_p$, where $\omega_p$ is plasma frequency, we will keep the time-integral in the collisional term of the kinetic equation.

### 3.2. Correlation tensors of the second order

In our model we will assume that the stochastic electromagnetic field that accelerates the charged particle is a random electromagnetic one and the charged particles have large and closed velocities to the velocity of the light. We also emphasize that the distribution of the magnetic inhomogeneities is anisotropic. This is the case of the regime of the chaotic acceleration occurring in gun effect when resonances overlap. The statistical properties of the random field might be described with a (infinite) sequence of the multi-point correlation functions, the most important of which is the (two-point) second-order correlation function. We will emphasize that the spectrum of the inhomogeneities is power law distribution. The correlation tensor of the second order that may describe a power law distribution of the inhomogeneities of the magnetic field and also a random behavior could be of the form

$$K_{\alpha\beta}(\mathbf{k}) = \frac{2 \langle E_{st}^2 \rangle k_{\min,y}^{s_y+1} k_{\min,\perp y}^{s_{\perp y}+1} \Gamma\left(\frac{s_y}{2}+1\right)\Gamma\left(\frac{s_{\perp y}}{2}+1\right)}{\left(k_{\min,y}^2 + k_{y,\alpha}^2\right)^{s_y/2+1} \left(k_{\min,\perp y}^2 + k_{\perp y,\beta}^2\right)^{s_{\perp y}/2+1} \Gamma\left(\frac{s_y}{2}-\frac{1}{2}\right)\Gamma\left(\frac{s_{\perp y}}{2}-\frac{1}{2}\right)} \qquad (3.2.1)$$

where $s_y$ and $s_{\perp y}$ are the spectral indices toward y axis and any other direction perpendicular to y axis, respectively, $s_y \neq s_{\perp y}$, $\Gamma(a)$ is gamma function. Here we adopt the y axis having the same direction of the y axis mentioned in the paper [15]. Hence the "$\perp y$" direction is an arbitrary direction perpendicular to y axis. Equation (3.2.1) has the following properties: (i) when $k_y$ (or $k_{\perp y}$) $\rightarrow 0$ while $k_{\perp y}$ (or $k_y$) remains constant, then $K_{\alpha\beta}$ (k) ~ const.(ii) when $k_y$ (or $k_{\perp y}$) $\rightarrow \infty$ while $k_{\perp y}$ (or $k_y$) remains constant, then $K_{\alpha\beta}$ (k) $\propto k_y^{\frac{1}{2}-s_y}$ ( or $K_{\alpha\beta}$ (k) $\propto k_{\perp y}^{\frac{1}{2}-s_{\perp y}}$ ), i.e., the scaling in both y and $\perp$ y directions are not the same. On the other hand one must point out that we will use the correlation tensor for $k_{min} << k << k_{max}$. For instance, numerical experiments showed strong anisotropy for $k_{min\, y}^{-1} = 100, k_{min\perp y}^{-1} = 400, s_y = 0.4, s_{\perp y} = 0.75$ [17].

While the geometry of the electromagnetic field, adopted by Argyris and Ciubotariu[15] is strictly directed to the Oy and Oz axis, we assumed in the above form of the correlation tensor a more realistic geometry, i.e. an anisotropic distribution. For further derivations we will suppose an appropriate choice of the correlation lengths and spectral indices corresponding to large anisotropy toward z axis for the magnetic field $B_b$ and y axis for the electric field $E_b$

### 3.3. Solution for the kinetic equation

In what it follows we will take account of the assumptions adopted for the equation (3.1.10). It is readily to express the equation (3.2.1) into another form if we take into account that the averaged right hand of the equation mentioned above is obtained from the term $\mathbf{E}\dfrac{d}{d\mathbf{p}}$ contained in the equation (3.1.1).

$$<\widehat{M}f^{(1)}> = e^2 \int d\tau E_{by\alpha} \frac{d}{dp_{y\alpha}} E_{by\beta} \frac{df}{dp_{y\beta}} + e^2 \int d\tau E_{by\alpha} \frac{d}{dp_{y\alpha}} E_{bx\beta} \frac{df}{dp_{x\beta}} +$$

$$e^2 \int d\tau E_{bx\alpha} \frac{d}{dp_{x\alpha}} E_{by\beta} \frac{df}{dp_{y\beta}} + e^2 \int d\tau E_{bx\alpha} \frac{d}{dp_{x\alpha}} E_{bx\beta} \frac{df}{dp_{x\beta}} = e^2 \int d\tau \frac{d}{dp_{y,\alpha}} K_{\alpha\beta,y} \frac{df}{dp_{y,\beta}} +$$

$$e^2 \int d\tau \frac{d}{dp_{y,\alpha}} K_{\alpha\beta} \frac{df}{dp_{\perp y,\beta}} + e^2 \int d\tau \frac{d}{dp_{\perp y,\alpha}} K_{\alpha\beta} \frac{df}{dp_{y,\beta}} + e^2 \int d\tau \frac{d}{dp_{\perp y,\alpha}} K_{\alpha\beta\perp y} \frac{df}{dp_{\perp y,\beta}} \quad (3.3.1)$$

Here $<\widehat{M}f^{(1)}>$ is the collision time integral of the equation (3.1.10) and

$$K_{\alpha\beta,y}(k) = \frac{2\langle E_{st}^2 \rangle k_{min,y}^{2(s_y+1)} \Gamma^2\left(\dfrac{s_y}{2}+1\right)}{\left(k_{min,y}^2 + k_{y,\alpha}^2\right)^{s_y/2+1} \left(k_{min,y}^2 + k_{y,\beta}^2\right)^{s_y/2+1} \Gamma^2\left(\dfrac{s_y}{2}-\dfrac{1}{2}\right)} \quad (3.3.2)$$

$$K_{\alpha\beta,\perp y}(k) = \frac{2\langle E_{st}^2 \rangle k_{min,\perp y}^{2(s_{\perp y}+1)} \Gamma^2\left(\dfrac{s_{\perp y}}{2}+1\right)}{\left(k_{min,\perp y}^2 + k_{\perp y,\alpha}^2\right)^{s_{\perp y}/2+1} \left(k_{min,\perp y}^2 + k_{\perp y,\beta}^2\right)^{s_{\perp y}/2+1} \Gamma^2\left(\dfrac{s_{\perp y}}{2}-\dfrac{1}{2}\right)} \quad (3.3.3)$$

are the correlation tensors adopted in the equation (3.3.1). In order to investigate how the operator $\frac{\partial}{\partial \mathbf{p}}$ acts onto the correlation tensor one must express the correlation tensor by its Fourier transforms

$$K_{\alpha\beta}(\mathbf{r},t;\Delta\mathbf{r}) = \int_{-\infty}^{+\infty} d\mathbf{k}\, e^{i\mathbf{k}\Delta\mathbf{r}} K_{\alpha\beta}(\mathbf{k}) \qquad (3.3.4)$$

where $\Delta \mathbf{r} = \mathbf{v}\tau$.

Using the Fourier transforms of the correlation tensors, the collision time integral turns out to be

$$<\hat{M}f^{(1)}> = \frac{2^{-s_y+1} \pi e^2 c^4 \langle E_{st}^2 \rangle i k_y \Gamma\left(\frac{s_y}{2}+1\right)}{\varepsilon^2 \Gamma\left(\frac{s_y}{2}-\frac{1}{2}\right)\Gamma\left(s_y+\frac{1}{2}\right)} [k_{min,y} v_y \tau]^{s_y} K_{s_y}[k_{min,y} v_y \tau] \times$$

$$\left\{ \frac{2^{-s_y} k_{min,y} \Gamma\left(\frac{s_y}{2}+1\right)}{\Gamma\left(\frac{s_y-1}{2}\right)\Gamma\left(s_y+\frac{1}{2}\right)} \int \tau^2 d\tau \, [2s_y a^{s_y-1} K_{s_y}(a) - a^{s_y} K_{s_y+1}(a)] \, f + \right.$$

$$\frac{2^{-s_{\perp y}} k_{min,\perp y} \Gamma\left(\frac{s_{\perp y}}{2}+1\right)}{\Gamma\left(\frac{s_{\perp y}-1}{2}\right)\Gamma\left(s_{\perp y}+\frac{1}{2}\right)} \int \tau^2 d\tau \, [2s_{\perp y} b^{s_{\perp y}-1} K_{s_{\perp y}}(b) - b^{s_{\perp y}} K_{s_{\perp y}+1}(b)] \, f \Big\} +$$

$$\frac{2^{-s_{\perp y}+1} \pi e^2 c^4 \langle E_{st}^2 \rangle i k_{\perp y} \Gamma\left(\frac{s_{\perp y}}{2}+1\right)}{\varepsilon^2 \Gamma\left(\frac{s_{\perp y}}{2}-\frac{1}{2}\right)\Gamma\left(s_{\perp y}+\frac{1}{2}\right)} [k_{min,\perp y} v_{\perp y} \tau]^{s_{\perp y}} K_{s_{\perp y}}[k_{min,\perp y} v_{\perp y} \tau] \times$$

$$\left\{ \frac{2^{-s_{\perp y}} k_{min,\perp y} \Gamma\left(\frac{s_{\perp y}}{2}+1\right)}{\Gamma\left(\frac{s_{\perp y}-1}{2}\right)\Gamma\left(s_{\perp y}+\frac{1}{2}\right)} \int \tau^2 d\tau \, [2s_{\perp y} b^{s_{\perp y}-1} K_{s_{\perp y}}(b) - b^{s_{\perp y}} K_{s_{\perp y}+1}(b)] \, f + \right.$$

$$\frac{2^{-s_y} k_{min,y} \Gamma\left(\frac{s_y}{2}+1\right)}{\Gamma\left(\frac{s_y-1}{2}\right)\Gamma\left(s_y+\frac{1}{2}\right)} \int \tau^2 d\tau \, [2s_y a^{s_y-1} K_{s_y}(a) - a^{s_y} K_{s_y+1}(a)] \, f \Big\} \qquad (3.3.5)$$

where $a = k_{min,y} v_y \tau$ and $b = k_{min,\perp y} v_{\perp y} \tau$. One can see from the above equation that for different magnitudes of the correlation lengths and spectral indices of the magnetic (electric) field along y, respectively $\perp$y directions the two terms contained in the equation (3.3.5) have different magnitudes. For instance if we adopt $s_y = 0.5$, $s_{\perp y} = 1.5$, and $k_{min\,y}^{-1} = 0.25 k_{min\perp y}^{-1}$ then $<\widehat{Mf}^{(1)}>_{\perp y} / <\widehat{Mf}^{(1)}>_y \leq 0{,}182$, where $<\widehat{Mf}^{(1)}>_i$ (i=y, $\perp$ y ) are the two terms of the equation(3.3.5), each one being proportional to $k_y$, respectively to $k_{\perp y}$. The distribution function $f$ is approximately constant during the correlation time $\tau$.

This ratio depends on the Lorentz factor as

$$<\widehat{Mf}^{(1)}>_{\perp y} / <\widehat{Mf}^{(1)}>_y = 0.494 \times \exp(\frac{\sqrt{1-\gamma^{-2}} - 0.25 c^{-1} v_{\perp y} - 0.75}{0.25 c^{-1} v_{\perp y} - 0.25}) \quad (3.3.6)$$

where this form was inferred for $s_y = 0.5$, $s_{\perp y} = 1.5$, and therefore depends also on the spectral indices of the electromagnetic field along two directions. It is worth noting that the collision time integral $<\widehat{Mf}^{(1)}>$ depends exponentially on the product $k_{min} v \tau$.

If in the equation (3.1.10) we replace the function $\tilde{f}$ (**r**-$\Delta$ **r**($\tau$ ),**p**, $t - \tau$ ) by its Fourier transform then the equation (3.1.10) becomes

$$-i(\omega - \mathbf{kv}) f_{\mathbf{k},\omega} = <\widehat{Mf}^{(1)}>_{\mathbf{k},\omega} \quad (3.3.7)$$

where $<\widehat{Mf}^{(1)}>_{\mathbf{k},\omega}$ stands for the collision time integral where we replace $f$ (**r**-$\Delta$ **r**($\tau$ ),**p**-$\Delta$ **p**($\tau$ ), $t - \tau$ ) by its Fourier transform $f_{\mathbf{k},\omega}$. It is worth noting that in the equation (3.3.7) frequency $\omega$ and wave-vector **k** correspond to a radiating wave. But the above equation (3.3.7) may also be written as

$$\frac{\partial f_{\mathbf{k},\omega}}{\partial \tau} + i \mathbf{kv}\, f_{\mathbf{k},\omega} = <\widehat{Mf}^{(1)}>_{\mathbf{k},\omega} \quad (3.3.8)$$

Substituting in the equation (3.3.8) the distribution function in the form

$$f_k(\tau) = v_\alpha^{-1} v_\beta^{-1} \delta(v_\alpha - v_{0\alpha}) \delta(v_\beta - v_{0\beta}) \exp(-i\omega\tau) \times g(\tau) \quad (3.3.9)$$

one leads to the equation $\quad \frac{\partial g}{\partial \tau} - i\omega g + i\mathbf{kv}g = <\widehat{Mf}^{(1)}>_{\mathbf{k},\omega,g} \quad (3.3.10)$

where $<\widehat{Mf}^{(1)}>_{\mathbf{k},\omega,g}$ stands for the collision time integral where we replace $f_{\mathbf{k},\omega}$ by $g(\tau)$.

The solution of the above equation (3.3.10) is of the form

$$g = g_0 \exp i \int d\tau \{\omega - \mathbf{kv} + <\widehat{Mf}^{(1)} / i>_{\mathbf{k},\omega,1}\} \quad (3.3.11)$$

with the initial condition

$$g_0 = \delta(\tau) \quad (3.3.12)$$

where for $<\widehat{Mf}^{(1)} / i>_{\mathbf{k},\omega,1}$ we formally replaced $f_{\mathbf{k},\omega}$ by unit 1.

Using equations (3.3.9), (3.3.11) and (3.3.12) the probability $f(\tau)$ may be written as

$$f_\mathbf{k}(\tau) = v_\alpha^{-1} v_\beta^{-1} \delta(v_\alpha - v_{0\alpha}) \delta(v_\beta - v_{0\beta}) \exp(-i\omega\tau) \times \exp i\int d\tau \{\omega - \mathbf{kv} + <\widehat{Mf}^{(1)} / i>_{\mathbf{k},\omega,1}\}$$
(3.3.13)

For small angles between **k** and **v** one can approximate, which takes place at high energy,

$$\omega - \mathbf{kv} \cong \frac{\omega}{2\gamma^2} \qquad (3.3.14)$$

where $\gamma$ is the Lorentz factor of the particle. On the other hand, from the equation (3.3.5) we will adopt the following notation

$$<Nf^{(1)}/i>_{k,\omega,1} = \frac{\varepsilon^2 <\widehat{M}f^{(1)}/i>_{k,\omega,1}}{\omega_{st}} \qquad (3.3.15)$$

Therefore the distribution function from (3.3.13) yields

$$f_k(\tau) = v_\alpha^{-1} v_\beta^{-1} \delta(v_\alpha - v_{0\alpha})\delta(v_\beta - v_{0\beta}) \exp(-i\omega\tau) \times \exp i\int d\tau \frac{\omega}{\gamma^2}\{\frac{1}{2} + \frac{\omega_{st}}{\omega}[<\widehat{N}f^{(1)}/i>_{k,\omega,1}$$

$$]/m_0^2 c^4\} \qquad (3.3.16)$$

The above distribution function (3.3.13) describes instability $\dfrac{\partial(\text{Re} f_k)}{\partial \gamma} < 0$ in the choasons' flow for energies which satisfy

$$\cos[-\omega\tau + \int \frac{\omega}{\gamma^2}(0.5 + \frac{\omega_{st}}{\omega} \cdot \frac{<\widehat{N}f^{(1)}/i>_{k,\omega,1}}{m_0^2 c^4})d\tau < (\gamma^{-2} - 1)[\int \omega(0.5 + \frac{\omega_{st}}{\omega} \cdot \frac{<\widehat{N}f^{(1)}/i>_{k,\omega,1}}{m_0^2 c^4})d\tau] \times$$

$$\sin[-\omega\tau + \int \frac{\omega}{\gamma^2}(0.5 + \frac{\omega_{st}}{\omega} \cdot \frac{<\widehat{N}f^{(1)}/i>_{k,\omega,1}}{m_0^2 c^4})d\tau \qquad (3.3.17)$$

In what it follows we will discuss the above inequality assuming that $\omega, \gamma$ and $<\widehat{N}f^{(1)}/i>_{k,\omega,1}$ do not change during time correlation $\tau$ and we will adopt $\tau = \omega^{-1} = \omega_{st}^{-1}$ where

$$\omega_{st} = \frac{e <B_b^2>^{1/2}}{m_0 c} \qquad (3.3.18)$$

is the stochastic gyrofrequency in the stochastic magnetic field $B_b$. On the other hand we assume that $\omega_{st}$ satisfies the equation (2.4). Therefore the instability condition may be put in the form

$$\cos[-1 + \gamma^{-2} \cdot (0.5 + \eta \cdot \frac{<\widehat{N}f^{(1)}/i>_{k,\omega,1}}{m_0^2 c^4})] < (\gamma^{-2} - 1)(0.5 + \eta \cdot \frac{<\widehat{N}f^{(1)}/i>_{k,\omega,1}}{m_0^2 c^4}) \times$$

$$\sin[-1 + \gamma^{-2} \cdot (0.5 + \eta \cdot \frac{<\widehat{N}f^{(1)}/i>_{k,\omega,1}}{m_0^2 c^4})] \qquad (3.3.19)$$

In the above inequation we labeled $\eta = \dfrac{\omega_{st}}{\omega}$ the ratio of the stochastic frequency to the frequency of the particle emission. That is, for instance we adopt $\omega_{st} \neq \omega$.

The figure 3.1 shows as the particle energy $\gamma$ increases the "turbulence energy" $[\dfrac{<\widehat{N}f^{(1)}/i>_{k,\omega,1}}{m_0^2 c^4}]^{1/2}$, achieved during chaotic acceleration, increases too. At a given energy of the particle there are regions of turbulence energy allowed by the above inequation (3.3.19) but also some other not allowed. Conversely at a given turbulence energy there are regions of particle energy allowed and other ones not allowed. The choice $\frac{\omega_{st}}{\omega} = 1$ corresponds to the "power law" region of the spectrum of the synchrotron radiation. For high ratio $\eta$ the distribution function $f_k(\tau)$ supports the instability condition wide spread magnitudes of the particle energy and turbulence energy.

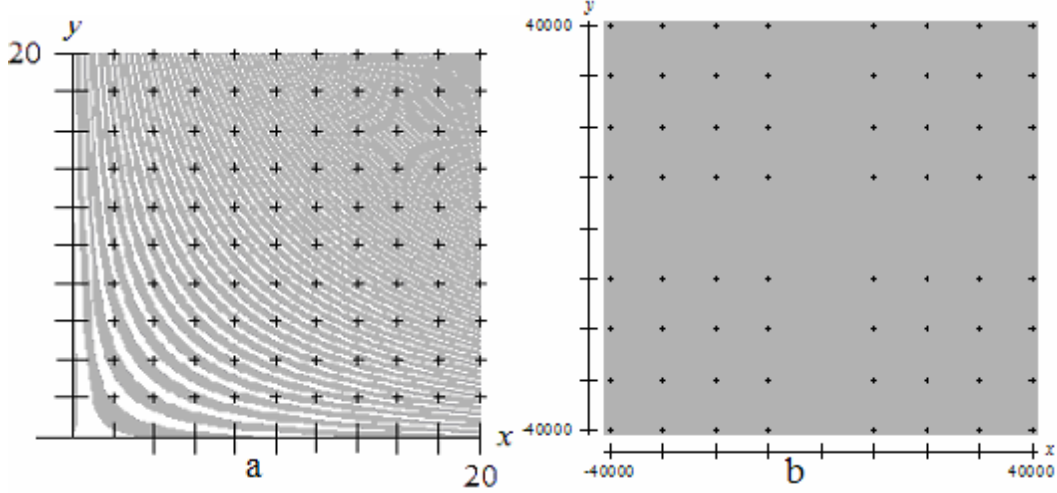

Fig.3.1 The plot corresponding to the inequation (3.3.19). The dashed domains are those where the inequality works; the x axis stands for $\gamma^{-2}$ while the y axis stands for $\frac{<\widehat{N}f^{(1)}/i>_{k,\omega,1}}{m_0^2 c^4}$ and $\frac{\omega_{st}}{\omega} = 1$ (a)

and $\frac{\omega_{st}}{\omega} = 10^3$ (b)

## 4. The spectrum emitted by charged particles in chaotic gun effect

An energy $\mathcal{E}_{n,\omega}$ radiated by the particle in a direction of **n** at a frequency $\omega$ is found as a flux of Poynting vector within a solid angle. If the Fourier component of the radiation field generated by one particle is given by [3]

$$\mathbf{B}_{n,\omega} = \frac{ie}{2\pi c^2 R} e^{ikR} \int_{-\infty}^{+\infty} [\mathbf{k},\mathbf{v}(t)] e^{i[\omega t - \mathbf{kr}(t)]} dt \qquad (4.1)$$

where $e$, $\mathbf{r}(t)$ and $\mathbf{v}(t)$ are the particle charge, radius-vector, and velocity, respectively; **n** is a unit vector along a line of sight, and **k** is the wave-vector of the radiating electro-magnetic wave in a medium. Then the energy $\mathcal{E}_{n,\omega}$ radiated by the particle in a direction **n** at a frequency $\omega$ is of the form

$$\mathcal{E}_{n,\omega} =$$

$$= \frac{e^2 \omega^2}{2\pi^2 c^3} \operatorname*{Re}_{T\to\infty} \int_{-T}^{T} dt \int_{0}^{\infty} d\tau e^{i\omega\tau} \int e^{-i\omega\tau} d^3v \cdot d^3v' \cdot d^3r \cdot d^3r' \; [\mathbf{n},\mathbf{v'}][\mathbf{n},\mathbf{v}] \, F(\mathbf{v},t) \, f(\mathbf{v},\mathbf{v'},\tau) \qquad (4.2)$$

where $f$ is the probability that a particle in a state (**r**,**v**) at a moment $t$ will appear in a state (**r'**,**v'**) at a moment $t+\tau$. $F$ (**r**,**v**, $t$) is the particle distribution function. The initial condition is described by

$$f(\mathbf{v},\mathbf{v'},0) = \delta(\mathbf{v}_\alpha - \mathbf{v}_\beta) \qquad (4.3)$$

$$F(\mathbf{v},0) = \delta(\mathbf{v}-\mathbf{v_0}) \qquad (4.4)$$

To find an energy radiated per unit time (i.e., the radiation intensity $I_{\mathbf{n},\omega}$) one must divide $\mathcal{E}_{\mathbf{n},\omega}$ by the total time $2T$. This is equivalent to omitting the integration over $dt$.

After the integration of the equation (4.2) over $v_0$ the spectral intensity is

$$I_\omega = \frac{e^2 \omega \varepsilon^2}{2\pi^2 c^3} \times \frac{\sin \int \frac{d\tau \cdot \omega}{\varepsilon^2}[\frac{m_0^2 c^4}{2} + \eta < Nf^{(1)}/i >_{k,\omega,1}]}{\frac{m_0^2 c^4}{2} + \eta < Nf^{(1)}/i >_{k,\omega,1}} \qquad (4.5)$$

In (4.5) we normalized the distribution function $F$ (**r**,**v**,t) and we took account that $\varepsilon$ and $\omega$ vary slowly during the correlation time.

The power emitted by an ensemble of charged particles can be derived using the following formula

$$P(\omega) = \int_{\varepsilon_1}^{\varepsilon_2} I_\omega(\varepsilon) dN(\varepsilon) \qquad (4.6)$$

For energies $\varepsilon_1 < \varepsilon < \varepsilon_2$ of the charged particle we will assume a power law dependence of the number of the particles

$$dN(\varepsilon) = K_e \varepsilon^{-\zeta} d\varepsilon \qquad (4.7)$$

where $\zeta$ is the spectral index.

Hence substituting $dN$ from equation (4.6) and $I_\omega$ from equation (4.7) the power yields

$$P(\omega) = \frac{e^2 \omega^2}{2\pi^2 c^3} K_e \int_{\varepsilon_1}^{\varepsilon_2} \varepsilon^{-\zeta} d\varepsilon \times \cos \int d\tau [\omega - \mathbf{k}\mathbf{v} + < \widehat{M} f^{(1)}/i >_{k,\omega,1}] \qquad (4.8)$$

and also

$$P(\omega) = \frac{e^2 \omega}{2\pi^2 c^3} K_e \int_{\varepsilon_1}^{\varepsilon_2} \varepsilon^{-\zeta} d\varepsilon \times \frac{\sin \int d\tau \frac{\omega}{\varepsilon^2}[\frac{m_0^2 c^4}{2} + \eta < Nf^{(1)}/i >_{k,\omega,1}]}{\frac{1}{\varepsilon^2}[\frac{m_0^2 c^4}{2} + \eta < Nf^{(1)}/i >_{k,\omega,1}]} \qquad (4.9)$$

For arbitrary index $\zeta$ the above integral cannot be performed analytically, but for natural numbers.

$$P(\omega) = \frac{-e^2\omega K_e}{4\pi^2 c^3 [\frac{m_0^2 c^4}{2} + \eta <Nf^{(1)}/i>_{k,\omega,1}]} \{ -\frac{\varepsilon^{-(\zeta-5)} \cos \int d\tau\omega\varepsilon^{-2}(\frac{m_0^2 c^4}{2} + \eta <Nf^{(1)}/i>_{k,\omega,1})]}{\int d\tau\omega[\frac{m_0^2 c^4}{2} + \eta <Nf^{(1)}/i>_{k,\omega,1}]} +$$

$$\frac{(\frac{\zeta-5}{2})\varepsilon^{-(\zeta-7)} \sin \int d\tau\omega\varepsilon^{-2}(\frac{m_0^2 c^4}{2} + \eta <Nf^{(1)}/i>_{k,\omega,1})]}{[\int d\tau\omega(\frac{m_0^2 c^4}{2} + \eta <Nf^{(1)}/i>_{k,\omega,1})]^2} -$$

$$\frac{(\frac{\zeta-5}{2})(\frac{\zeta-7}{2})}{[\int d\tau\omega(\frac{m_0^2 c^4}{2} + \eta <Nf^{(1)}/i>_{k,\omega,1})]^2} \int \varepsilon^{-(\zeta-9)} \sin \int d\tau\omega\varepsilon^{-2}(\frac{m_0^2 c^4}{2} + \eta <Nf^{(1)}/i>_{k,\omega,1})] d(\varepsilon^{-2}) \}$$

(4.10)

If the spectral index of the number of the charged particles is $\zeta = 7$ then an analytical form of the power is

$$P(\omega) = \frac{e^2 K_e}{4\pi^2 c^3 m_0^2 c^4 (0.5 + \eta \frac{<Nf^{(1)}/i>_{k,\omega,1}}{m_0^2 c^4})} \cdot \frac{\cos \int d\tau\omega\gamma^{-2}(0.5 + \eta \frac{<Nf^{(1)}/i>_{k,\omega,1}}{m_0^2 c^4})}{\tau(0.5 + \eta \frac{<Nf^{(1)}/i>_{k,\omega,1}}{m_0^2 c^4})} -$$

$$-\frac{e^2 K_e}{4\pi^2 c^3 m_0^2 c^4 (0.5 + \eta \frac{<Nf^{(1)}/i>_{k,\omega,1}}{m_0^2 c^4})} \frac{\sin \int d\tau\omega\gamma^{-2}(0.5 + \eta \frac{<Nf^{(1)}/i>_{k,\omega,1}}{m_0^2 c^4})}{\omega\tau^2 m_0^4 c^8 (0.5 + \eta \frac{<Nf^{(1)}/i>_{k,\omega,1}}{m_0^2 c^4})^2} \quad (4.11)$$

In this derivation (4.11) we take account that during correlation time $\tau$ the energy and the frequency vary very slowly.

The power derived in (4.11) has fluctuations along the frequency variation. The peaks of this fluctuations lie along a line which describes a power law of the spectrum when one assumes $\eta = 1$.

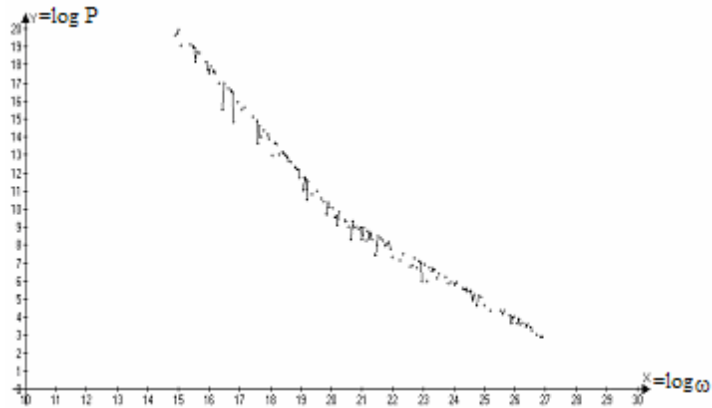

Fig.4.1 The power spectrum of the synchrotron emission derived in (4.10) where log P is normalized to $e^2 K_e (2\pi^2 c^3)^{-1} \tau^{-1} [0.5(m_0 c^2)^2 + <Nf(1)>]^{-1}$

The power law $P(\omega) \sim \omega^{-\alpha}$ obtained for the spectral index $\zeta = 9$ of the number of particles leads to α=2.0698965 for x < 20 and α= 1.0142343 for x > 20
On the other hand, as one can see in Fig.4.2, the slope of spectrum depends on the spectral index of the number of the charged particles.

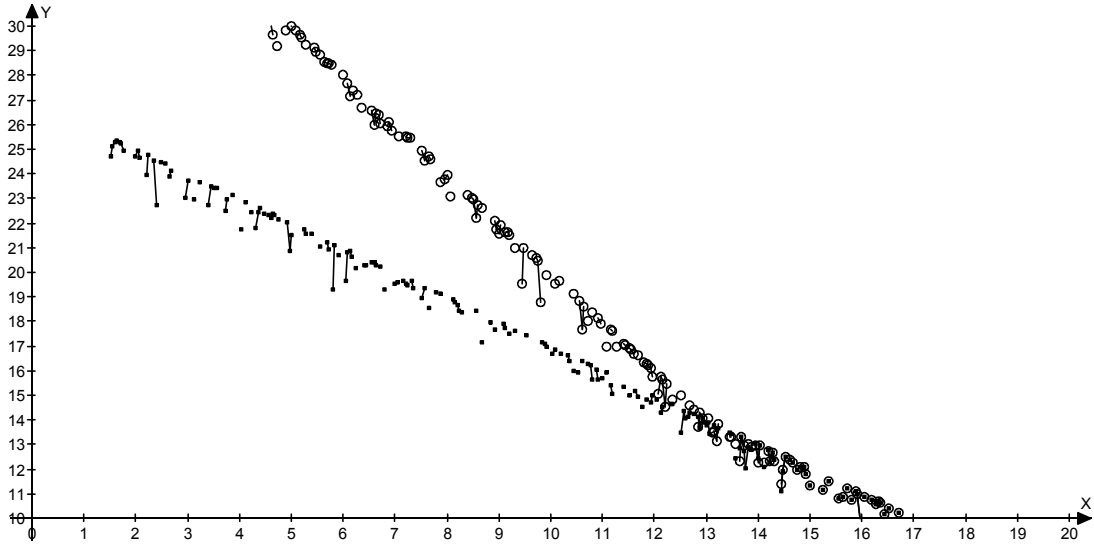

Fig.4.2. Two power spectra of the synchrotron emission for spectral indices of the number of particles $\zeta = 7$ (full squares) and $\zeta = 9$ (empty circles). Y stands for log $P(\omega)$ and x stands for log $\omega$; both are in arbitrary units

## 5.Astrophysical consequences

Numerical experiment performed by Argyris and Ciubotariu provide a proof that a random electromagnetic field with small-scale inhomogeneities and special geometry can accelerate charged particles even from low energies to very high ones. This occurs when the frequency of the inhomogeneities are much larger than the frequency of large-scale magnetic field.

Their work doesn't assume any cut-off in the energy spectrum. In our model the cut-off is constrained by the scale length of the source where the ratio of the characteristic frequency of the emission of the charged particles and that of the large-scale field meet the requirement $\beta \geq 4$. Hence for instance, for radio source 3C 273 for which the projected length is about 39 kpc $\cong 39 \times 10^{21}$ cm one can estimate the cut-off of the energy to be $\varepsilon \cong eBr = 0.117 TeV$ for a magnetic strength of $10^{-5}G$ [18]. The numerical experiment of Argyris and Ciubotariu[15] leads to a magnitude of the characteristic frequency of inhomogeneities within a magnetic field in the range of 4 kHz for the large scale magnetic field mentioned above. 3C 273 is a radio-loud quasar, and was one of the first extragalactic X-ray sources discovered in 1970. However, even to this day, the process which gives rise to the X-ray emissions is controversial. Polarization with coincident orientation has been observed in radio, infrared, and optical light being emitted from the large-scale jet; these emissions are therefore almost certainly synchrotron in nature, a radiation that is created by a jet of charged particles moving at relativistic speeds.

Assuming a cylindrical geometry of the synchrotron emission with a subparsec diameter one may relate the power of the synchrotron emission $P(\omega)$ to its energetic flux $\Phi$ as

$$\Phi = \frac{P}{S} \quad (5.1)$$

where $S$ is the transverse area of the cylinder. For instance if we adopt $S \cong 10^{32} \, cm^2$ then for frequencies in the range $\omega \cong 10^{27} \, Hz$ we can obtain a power law form of the synchrotron emission power spectrum, i.e.

$$\log \frac{P}{S} = -\beta \log \omega \quad (5.2)$$

when

$$\log P = -\alpha \log \omega \quad (5.3)$$

if the index $\alpha$ of the power spectrum is

$$\alpha = \beta - 1.18 \quad (5.4)$$

Numerical simulations show that the differential spectrum obtained from equation (4.11) may be approached, for instance, by a function of the type

Y = log((10^(-2*x+80))*sin(10^(x-30))-(10^(-x+10))*cos(10^(x-30))-(10^3)*sin(10^(x-30)))    (5.5)

which could match the time-averaged energy spectrum of Mrk 501 in the interval from 1 TeV to 5 TeV [19]

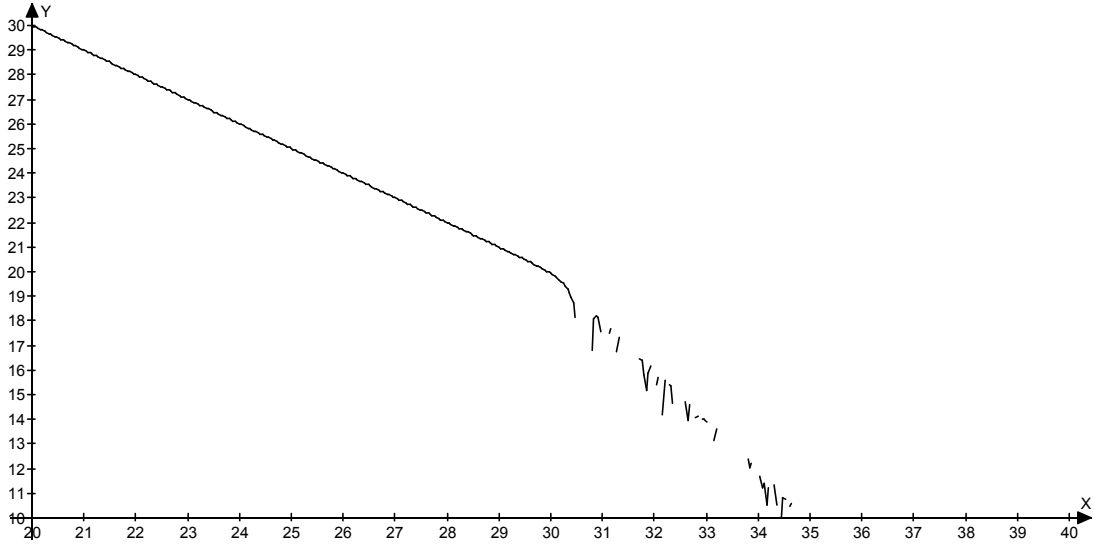

Fig.5.1. The plot of the function from (5.5) which can depicts the differential spectrum assigned to the synchrotron emission in arbitrary units

The function chosen above has a continuum shape up to x = 30, then the spectrum becomes discontinuous. For x< 30 the function plot may be fitted by linear function

$$Y = -1.0011276*x+50.0262051 \qquad (5.6)$$

The equation $\alpha = \beta - 1.18$, yields

$$\beta = 2.18112766 \qquad (5.7)$$

Aharonian et al.[19] found a differential index of -2.23 ±0.04$_{stat}$ and -2.26 ±0.06$_{stat}$ for the high and the low flux spectrum respectively in the energy region from 1 TeV to 5 TeV . The function adopted above predicts a break at approximately 15 TeV. Numerical simulations show that the break is related to the square of the ratio of the energy of the turbulent field (the stochastic field) to the particle's energy.

$$\omega_{break} = \log\{\tau \cdot \frac{0.5 + \frac{<Nf^{(1)}/i>_{k,\omega,1}}{m_0^2 c^4}}{\gamma^2}\}^{-1} \qquad (5.8)$$

Assuming $\omega_{break} \approx 10^{30} Hz$, $H = 0.2$ and $\Omega_B = 0.05$ one can estimate $B_0 = 10^{21} G$. Using the equations (5.8) and (3.3.19) , one can estimate the energy of the charged particle from the equation

$$\tau^2 - 0.5\gamma^{-2} = \frac{\frac{<Nf^{(1)}/i>_{k,\omega,1}}{m_0^2 c^4}}{\gamma^2} \qquad (5.9)$$

where we adopted $\tau^{-1} \approx \omega_{break}$. As one can see from the figure 3.1 one can adopt a different ratio $\frac{\frac{<Nf^{(1)}/i>_{k,\omega,1}}{m_0^2 c^4}}{\gamma^2}$ for which the chaotic gun effect could take place.

One can estimate the number density $n_b$ of the charged particles in the beam basing it on $\Omega_p = 0$ which we have adopted in our model. This would lead to $n_b \ll 3.14(kc)^2 \times 10^{-10} cm^{-3}$. For gamma rays in the TeV domain $(kc) \cong 10^{27} Hz$ it would yield $n_b \ll 10^{44}$ cm$^{-3}$.

Our model allows us to characterize the polarization of the radiation since the free path along the y direction are different from $\perp$ y direction(e.g., see Sect. 3.2). The polarization depends on the Lorentz factor, on the coherence lengths and on the spectral indices along those two directions. A detailed study of the polarization will be addressed in a future paper.

## 6. Discussions and conclusions

Assuming anisotropic correlation tensor of the second order for the random magnetic field, we derived a kinetic equation with a collision time integral that displays anisotropy of the path length of the charged particle. This leads to the anisotropy of the synchrotron emission of those particles involved in chaotic motion. Our model can support the numerical experiment of Argyris and Ciubotariu and provide a way to make estimations concerning the energy of the charged particle.

The square ratio of the energy of the turbulent field to the energy of the particle establishes the break of the spectrum between the continuum shape and the discontinuous one. It also establishes the break between the two spectral indices of the differential spectrum of the synchrotron emission which depict the shape mentioned above. The spectral index of the synchrotron emission spectrum is related to the spectral index of the number of the particles.

The distribution function of the charged particles inferred in section 3.3 can support continuum acceleration from low energies to high ones.

We performed in this paper an investigation of the spectrum only for natural numbers for the spectral index. A more detailed study can reveal other new features concerning the synchrotron emission.

### Acknowledgments

I'd like to thank Tim Bastian from National Radio Astronomy Observatory for his observations concerning my work and also Gelu Nitu from New Jersey Institute of Technology for the assistance he gave me in improving my reference list.